# Cyber-Physical Control over Wireless Sensor and Actuator Networks with Packet Loss

Feng Xia[*], Xiangjie Kong, and Zhenzhen Xu

*School of Software, Dalian University of Technology, Dalian 116620, China*

**Abstract.** There is a growing interest in design and implementation of cyber-physical control systems over wireless sensor and actuator networks (WSANs). Thanks to the use of wireless communications and distributed architectures, these systems encompass many advantages as compared to traditional networked control systems using hard wirelines. While WSANs are enabling a new generation of control systems, they also introduce considerable challenges for quality-of-service (QoS) provisioning. In this chapter we examine some of the major QoS challenges raised by WSANs, including resource constraints, platform heterogeneity, dynamic network topology, and mixed traffic. These challenges make it difficult to fulfill the requirements of cyber-physical control in terms of reliability and real-time. The focus of this chapter is on addressing the problem of network reliability. Specifically, we analyze the behavior of wireless channels via simulations based on a realistic link-layer model. Packet loss rate (PLR) is taken as a major metric for the analysis. The results confirm the unreliability of wireless communications and the uncertainty of packet loss over WSANs. To tackle packet loss, we present a simple solution that can take advantage of existing prediction algorithms. Simulations are conducted to evaluate the performance of several classical prediction algorithms used for packet loss compensation. The results give some insights into how to deal with packet loss in cyber-physical control systems over unreliable WSANs.

## 1. Introduction

A wireless sensor and actuator network (WSAN) [1-4] is a networked system of geographically distributed sensor and actuator nodes, as shown in Fig. 1. These nodes are interconnected via wireless links. The scale of the network depends highly on the target application. In general, both sensor and actuator nodes are equipped with some data processing and wireless communication capabilities, as well as power supply. Over the past years, a number of prototype and commercial wireless sensor nodes have been made available by research institutions and companies from around the world. Typical examples include BTnode, FireFly, IMote2, MicaZ, SunSPOT, TinyNode584, Tmote Sky, etc. Sensors gather information about the state of physical world and transmit the collected data to actuators through single-hop or multi-hop communications over the radio channel. Upon receipt of the required information, the actuators make the decision about how to react to this information and perform corresponding actions to change the behavior of the physical environment. As such, a closed loop is formed integrating the cyber and physical worlds. In addition to sensor and actuator nodes, there is commonly a base station in the WSAN, which is principally responsible for monitoring and managing the overall network through communicating with sensors and actuators.

---

[*] Corresponding Author; Email: f.xia@ieee.org.

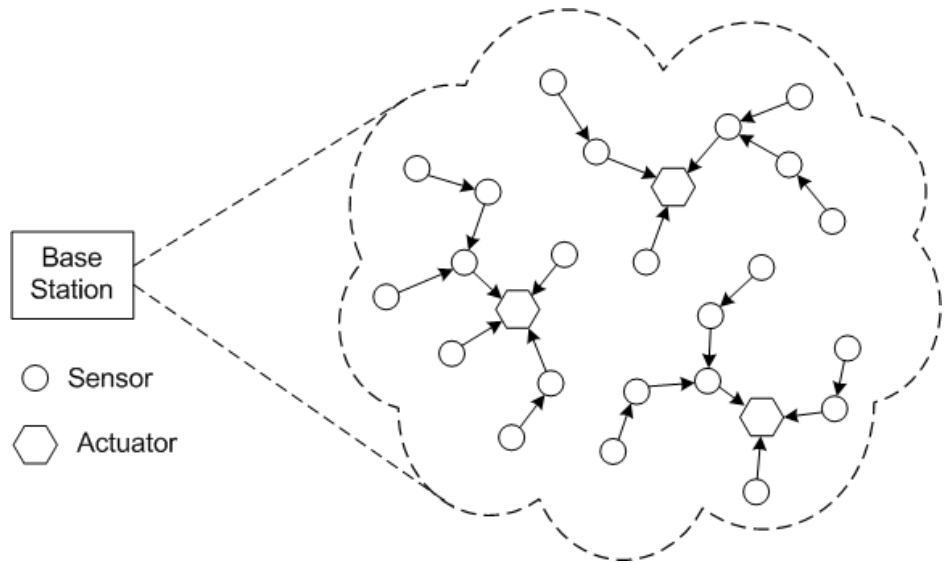

**Fig. 1.** A wireless sensor and actuator network

General wireless sensor networks (WSNs) are used for information gathering in applications like habitat monitoring, military surveillance, agriculture and environmental sensing, and health monitoring. The primary functionality of such WSNs is to sense and monitor the state of the physical world. In most cases, they are unable to *affect* the physical environment. In many applications, however, it is not sufficient to just observe the state of the physical system; it is also expected to respond to the sensed events/data by performing corresponding actions upon the physical system. For instance, in a fire handling system, it is necessary for the actuators to turn on the water sprinklers upon receipt of a report of fire. WSANs can satisfy such requirements by enabling the application systems to sense, interact, and change the physical world, e.g., to monitor and manipulate the lighting in a smart office or the speed of a mobile robot.

Through closing the loop involving both the cyber and physical worlds, WSANs will be one of the most critical technologies for building future cyber-physical systems (CPSs) [5,6], which promise to revolutionize the way we interact with the physical world. Such systems can be deployed in lots of applications such as health care, home automation, assisted living, intelligent building, intelligent transportation, disaster relief, planet exploration, and industrial control. In particular, WSANs exploit the methodology of feedback, which has been recognized as the central element of control systems. The advent of WSANs has the potential to revolutionarily promote existing control applications by enabling an unprecedented degree of distributed cyber-physical control.

Today's control systems are usually built upon hard wirelines. In contrast, control over WSANs exploits the potential of wireless communications, which deliver many advantages [7,8]. For instance, various difficulties related to the installation and maintenance of the large number of cables are completely eliminated. Consequently the flexibility and expandability of the system can be further enhanced. At the same time, system maintenance and update become easier, and the cost will of course be reduced. In some harsh environments it is forbidden or unfavorable to use cables due to constraints concerning e.g. physical environments and production conditions. This is especially the case when deleterious chemicals, severe vibrations and high temperatures are present that could potentially damage any sort of cabling. For such situations wireless technologies offer a much better choice for achieving connectivity. In addition, wireless control satisfies the requirements of mobile systems, enabling closed-loop control of mobile objectives such as automated guided vehicles, mobile robots, and unmanned aerial vehicles.

However, the use of wireless networks in connecting spatially distributed sensors, controllers, and actuators raises new challenges for control systems design [3,7]. Wireless channels have adverse properties such as path loss, multi-path fading, adjacent channel interference, Doppler shifts, and half-duplex operations. Consequently, transmitting radio signals over wireless channels can be affected by many factors, such as ambient noise, physical obstacles, node movement, environmental changes, and transmission power, to mention just a few. The inherent openness of wireless connections may potentially cause the operating environment of the system to be highly dynamic and unpredictable, since the wireless channel might be used by other co-existing devices. Wireless communications are much less dependable than wirelines in that the bit error rate of a wireless channel is usually several times that of a wired connection [9]. This is especially true for WSANs that feature low-power communications. As a consequence, constructing cyber-physical control systems over WSANs is challenging because the network quality-of-service (QoS) cannot always be guaranteed. Particularly, the control performance might be sacrificed due to unpredictable packet loss, which could even cause system instability in extreme cases.

This chapter aims to develop a better understanding of how to realize cyber-physical control over unreliable WSANs. For this purpose, major challenges with respect to QoS provisioning will be outlined. Special attention is given to the packet loss problem arising in the context of WSAN. Using a realistic link-layer model, we analyze the characteristics of the wireless channel in terms of packet loss rate (PLR). The results confirm the unreliability and uncertainty of wireless communications over WSANs. To cope with packet loss, we present a simple solution that can take advantage of existing prediction algorithms such as time series forecasting. Simulations are conducted to evaluate the performance of several classical prediction algorithms used for packet loss compensation.

The rest of the chapter is organized as follows. Section 2 summarizes some state-of-the-art work related to this chapter. In Section 3, we describe the architecture of cyber-physical control systems exploiting WSANs. Major QoS challenges will be discussed in Section 4. The behavior of wireless channel will be analyzed via simulations in Section 5, where we focus on PLR with respect to communication distance and transmission power. A simple approach to packet loss compensation is presented and evaluated in Sections 6 and 7 respectively. Section 8 concludes the chapter.

## 2. Related Work

Cyber-physical systems [5,10] are integrations of computation, networking, and physical dynamics, in which embedded devices such as sensors and actuators are (wirelessly) networked to sense, monitor and control the physical world. It is believed in both the academic and industrial communities that CPS will have great technical, economic and societal impacts in the future. The CPS of tomorrow will far exceed those of today in terms of both performance and efficiency. The realm of CPS is opening up unprecedented opportunities for research and development in numerous disciplines, e.g. computing, communications, and control. In recent years, CPS has been attracting attention from a rapidly-increasing number of researchers and engineers. To fully exploit the potential of CPS, however, many challenges must be overcome.

Wireless sensor and actuator networks play an essential role in cyber-physical control systems, since they are the *bridge* between the cyber and physical worlds. Akyildiz and Kasimoglu [11] described research challenges for coordination and communication problems in WSANs. Melodia *et al.* [12] further studied these problems in WSANs with mobile actuators. Rezgui and Eltoweissy [13] discussed the opportunities and challenges for service-oriented sensor and actuator networks. Sikka *et al.* [14] deployed a large heterogeneous WSAN on a working farm to explore sensor network applications that can help manage large-scale farming systems. In comparison with the filed of general WSN in which significant progress has been made over the years, WSAN is a relatively new research area yet to be explored.

In particular, there is only limited work in the WSAN area targeting cyber-physical control applications. For example, Li [15] prototyped a light monitoring and control system as a case study of WSANs. Oh *et al.* [16] illustrated the main challenges in developing real-time control systems for pursuit-evasion games using a large-scale sensor network. Bosch *et al.* [17] reported the application of WSANs in distributed movement control and coordination of autonomous vehicles. Korber *et al.* [18] dealt with some of the design issues of a highly modular and scalable implementation of a WSAN for factory automation applications. Nikolakopoulos *et al.* [19] developed a gain scheduler for wirelessly networked control systems to cope with time-varying delay induced by multi-hop communications in sensor networks. Despite growing interest, the impact of packet loss, as a result of unreliable communications in WSANs, on the performance of cyber-physical control remains open for further research.

In the control community, significant effort has been made for packet loss compensation. A survey on this topic can be found in [20]. Despite their differences, most of existing packet loss compensation methods share the following features. First, they depend heavily on the knowledge about the accurate models of the physical systems to be controlled, and, possibly, the controller design. Second, the relevant algorithms are computationally intensive. For these reasons, they are impractical for real systems lacking well-established mathematical models. In addition, they are usually not desirable solutions for resource-constrained WSANs because of overly-large computational overheads.

This chapter is closely related to our previous work [2-4,6]. In [3], Xia *et al.* proposed an application-level design methodology for WSANs in mobile control applications. The unreliability of wireless links within sensor networks was also studied experimentally. QoS challenges and opportunities for WSANs were discussed in [2,6]. A QoS management scheme using fuzzy logic control technique and feedback scheduling concept was presented in [4,6].

## 3. System Architecture

Feedback cyber-physical control deals with the regulation of the characteristics of a cyber-physical system. The main idea of feedback control is to exploit measurements of the system's outputs to determine the control commands that yield the desired system behavior. As shown in Fig. 2, a controller, together with some sensors and actuators, is usually used to sense the operation of the physical system, compare it against the desired behavior, compute control commands, and perform actions onto the system to effect the desired change. This feedback architecture of a cyber-physical control system is also called *closed loop*, implying that the cyber space and the physical system are able to affect each other.

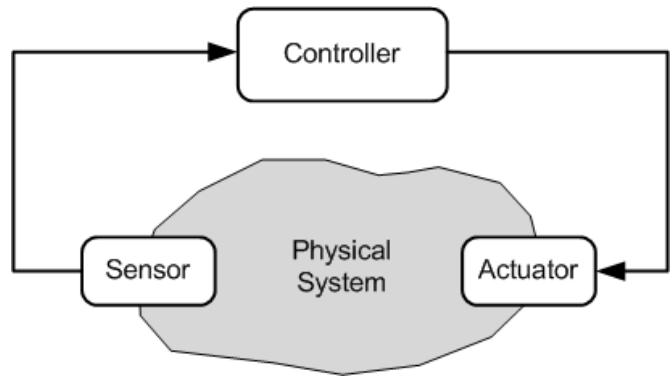

**Fig. 2.** A cyber-physical control system

As mentioned previously, there are three essential components in a WSAN: sensors, actuators, and a base station. Depending on whether there are explicit controller entities within the network, two types of system architectures of WSANs for cyber-physical control can be distinguished, as shown in Fig. 3 and Fig. 4 respectively [3]. These architectures are called automated architecture and semi-automated architecture respectively in [11].

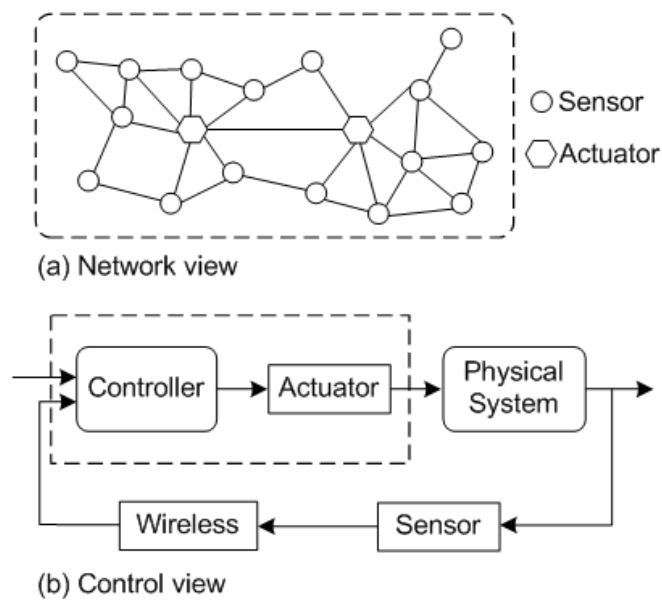

**Fig. 3.** WSAN Architecture without explicit controllers

In the first type of architecture as shown in Fig. 3(a), there is no explicit controller entity in the WSAN. In this case, controllers are embedded into the actuators and control algorithms for making decisions on what actions should be performed upon the physical systems will be executed on the actuator nodes. The data gathered by sensors will be transmitted directly to the corresponding actuators via single-hop or multi-hop communications. The actuators then process all incoming data by executing pre-designed control algorithms and perform appropriate actions. From the control perspective, the actuator nodes serve as not only the actuators but also the controllers in control loops. From a high-level view, wireless communications over WSANs are involved only in transmitting the sensed data from sensors to actuators. Control commands do not need to experience any wireless transmission because the controllers and the actuators are integrated, as shown in Fig. 3(b). In this chapter, we consider cyber-physical control systems with this architecture.

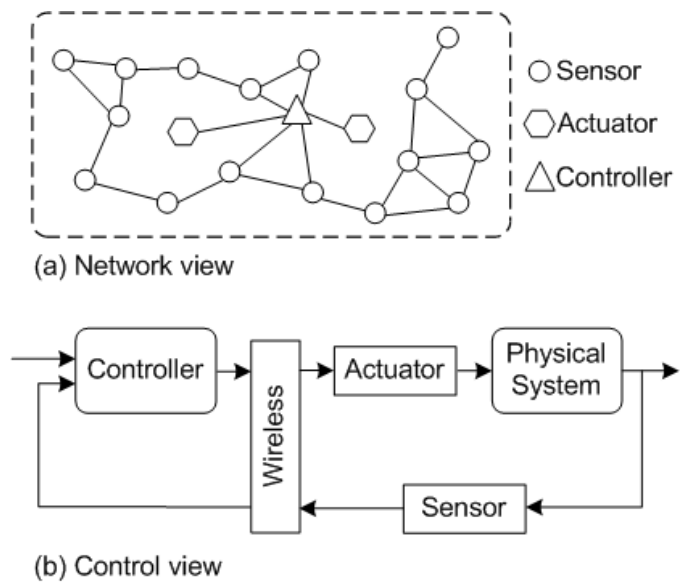

**Fig. 4.** WSAN Architecture with explicit controllers

Fig.4(a) shows an alternative type of architecture, in which one or more controller entities explicitly exist in the WSAN. The controller entities could be functional modules embedded in the base stations or separated nodes equipped with sufficient computation and communication capacities. With this architecture, sensors send the collected data to the controller entities. The controller entities then execute certain control algorithms to produce control commands and send them to actuators. Finally, the actuators perform the actions. In this context, both the sensor data and control commands need to be transmitted wirelessly in a single-hop or multi-hop fashion. A high-level view of the applications of this architecture is depicted in Fig. 4(b).

## 4. QoS Challenges

Cyber-physical control imposes considerable QoS requirements on WSANs. For instance, in a fire handling system built upon a WSAN, sensors need to report the occurrence of a fire to actuators in a timely and reliable fashion; then, the actuators equipped with water sprinklers will react by a certain deadline so that the situation will not become uncontrollable. Depending on the type of application, QoS in WSANs can be characterized by reliability, timeliness, robustness, availability, and security, among others.

WSAN in nature is a special category of wireless networks, which has its own characteristics besides the previously-mentioned properties of wireless channels. These unique characteristics make it quite difficult to provide QoS support in control systems over WSANs. Some major challenges in this context are described in the following [2].

1) Resource Constraints

Wireless sensor nodes are usually low-cost, low-power, small devices that are equipped with only limited data processing capability, transmission rate, battery energy, and memory. For example, the MICAz mote from Crossbow is based on the Atmel ATmega128L 8-bit microcontroller that provides only up to 8 MHz clock frequency, 128-KB flash program memory and 4-KB EEPROM; the transmit data rate is limited to 250 Kbps. Due to the limitation on transmission power, the available bandwidth and the radio range of the wireless channel are often limited. In particular, energy conservation is critically important for extending the lifetime of the network, because it is often infeasible or undesirable to recharge or replace the batteries attached to sensor nodes once they are deployed. Actuator nodes typically have stronger computation and communication capabilities and more energy budget relative to sensors. Despite this fact, resource constraints apply to both sensors and actuators.

In the presence of resource constraints, the network QoS may suffer from the unavailability of computing and/or communication resources. For instance, a number of nodes that want to transmit messages over the same WSAN have to compete for the limited bandwidth that the network is able to provide. As a consequence, some data transmissions will possibly experience large delays, resulting in low level of QoS. Due to the limited memory size, data packets may be dropped before the nodes successfully send them to the destination. Therefore, it is of critical importance to use the available resources in WSANs in a very efficient way.

2) Platform Heterogeneity

Sensors and actuators do not share the same level of resource constraints, as mentioned above. Possibly designed using different technologies and with different goals, they are different from each other in many aspects such as computing/communication capabilities, functionality, and number. In a large-scale system of systems, the hardware and networking technologies used in the underlying WSANs may differ from one subsystem to another. This is true because of the lack of relevant standards dedicated to WSANs and hence commercially available products often have disparate features. This platform heterogeneity makes it very difficult to make full use of the resources available in the integrated system. Consequently, resource efficiency cannot be maximized in many situations. In addition, the platform heterogeneity also makes it challenging to achieve real-time and reliable communication between different nodes.

3) Dynamic Network Topology

Unlike WSNs where (sensor) nodes are typically stationary, the actuators in WSANs may be mobile [12]. In fact, node mobility is an intrinsic nature of many applications such as, among others, intelligent transportation, assisted living, urban warfare, planetary exploration, and animal control. During runtime, new sensor/actuator nodes may be added; the state of a node is possibly changed to or from sleeping mode by the employed power management mechanism; some nodes may even die due to exhausted battery energy. All of these factors might potentially cause the network topologies of WSANs to change dynamically.

Dealing with the inherent dynamics of WSANs requires QoS mechanisms to work in dynamic and even unpredictable environments. In this context, QoS adaptation becomes necessary; that is, WSANs must be adaptive and flexible at runtime with respect to changes in available resources. For example, when an intermediate node dies, the network should still be able to guarantee real-time and reliable communication by exploiting appropriate protocols and algorithms.

4) Mixed Traffic

In many situations, diverse applications need to share the same WSAN, inducing both periodic and aperiodic data. This feature will become increasingly evident as the scale of WSANs grows. Some sensors may be used to create the measurements of certain physical variables in a periodic manner for the purpose of monitoring and/or control. Meanwhile, some others may be deployed to detect critical events. For instance, in a smart home, some sensors are used to sense the temperature and lighting, while some others are responsible for reporting events like the entering or leaving of a person. Furthermore, disparate sensors for different kinds of physical variables, e.g., temperature, humidity, location, and speed, generate traffic flows with different characteristics (e.g. message size and sampling rate). This feature of WSANs necessitates the support of service differentiation in QoS management.

## 5. Wireless Channel Characterization

In the following, we focus our attention on the problem of packet loss. In control systems, packet loss degrades control performance and even causes system instability. Because real-life control applications can only tolerate occasional packet losses with a certain upper bound of allowable packet loss rate, WSAN design should minimize the occurrence of packet losses as much as possible. Ideally, every packet should be transmitted successfully from the source to the destination without loss. However, due to many factors such as low-power radio communication, variable transmission power, multi-hop transmission, noise, radio interference, and node mobility, packet loss cannot be completely avoided in WSANs.

To elaborate on this issue, it is necessary to understand the characteristics of wireless channels used by WSANs. More specifically, we need to capture the wireless link quality in terms of packet loss rate. For this purpose, we perform simulations based on a realistic WSN link-layer model developed by Zuniga and Krishnamachari [21]. Some of the simulation results will be presented in this section. In the literature, experiments have been conducted for analysis of the link quality in WSNs, e.g., [22-25].

Simulation parameters used in the channel model are set according to the profile of MICA2 motes, as used in [21]. In this work, we examine the impact of two major factors:

communication distance (i.e. the physical distance between the transmitter and the receiver, denoted *d*) and transmission power (of the transmitter), though there are many other factors that could induce packet loss. For simplicity, the effect of medium access contention between different nodes is not taken into consideration.

Fig. 5 depicts the relationship between PLR and the transmitter-receiver distance. In this case, the transmission power is set to 0dBm (a high level). For every distance ranging from 1m to 15m with steps of 1m, 80 independent measures are given.

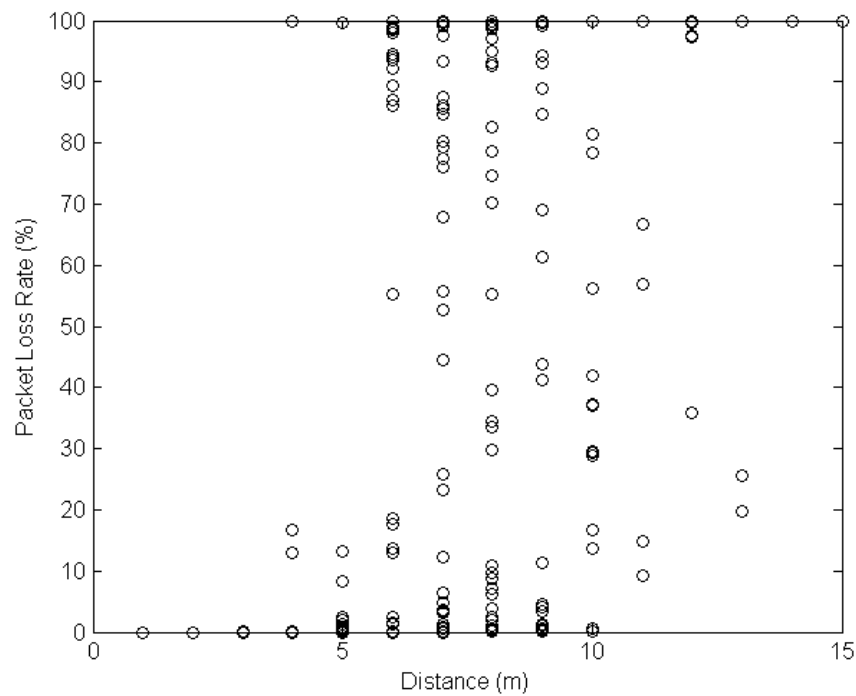

**Fig. 5.** Packet loss rate with respect to distance (transmission power: 0dBm)

As we can see from Fig.5, the link quality is highly related to the transmitter-receiver distance. Depending on the link quality, the whole area can be divided into three regions: connected region, transitional region, and disconnected region [21]. According to Fig.5, the connected region in this case corresponds to the distances between 0 and 3m. In this region, all packets sent by the transmitter will be received successfully by the destination node (i.e. the receiver), implying a PLR of zero. In contrast, when the receiver resides within the disconnected region (corresponding to $d \geq 14$m), no packet will be received. Approximately, the radio range is around 13m. These observations are very easy to understand since the strength of an electromagnetic signal decays with respect to distance during propagation. The packet cannot be received if the received signal strength is below the receive sensitivity of the destination node.

The transitional region deserves special attention, which corresponds to the distances between 4m and 13m. Within this region, the PLR associated with the link could vary drastically. This is true even for a given distance. For instance, for distances between 4m and 10m, the PLR could very between 0 and 100%. This demonstrates the uncertainty as well as unreliability of the wireless link. The uncertainty of PLR for a given distance can be explained by the fact that the signal strength is random and often log-normally distributed about the mean distance-dependent value [21].

To examine the influence of transmission power, we change the transmission power to a low level, i.e. -10dBm, and re-run the simulations. The results are reported in Fig. 6. It can be seen that the radio range is significantly reduced at a low level of transmission power. When the distance reaches 8m, the link becomes disconnected, with a PLR of 100%. In comparison with Fig.5, the size of the connected region shrinks in Fig.6. Within the transitional region ($3m \leq d \leq 7m$) in Fig.6, the PLR is highly uncertain, which is similar to the case of high transmission power (Fig.5).

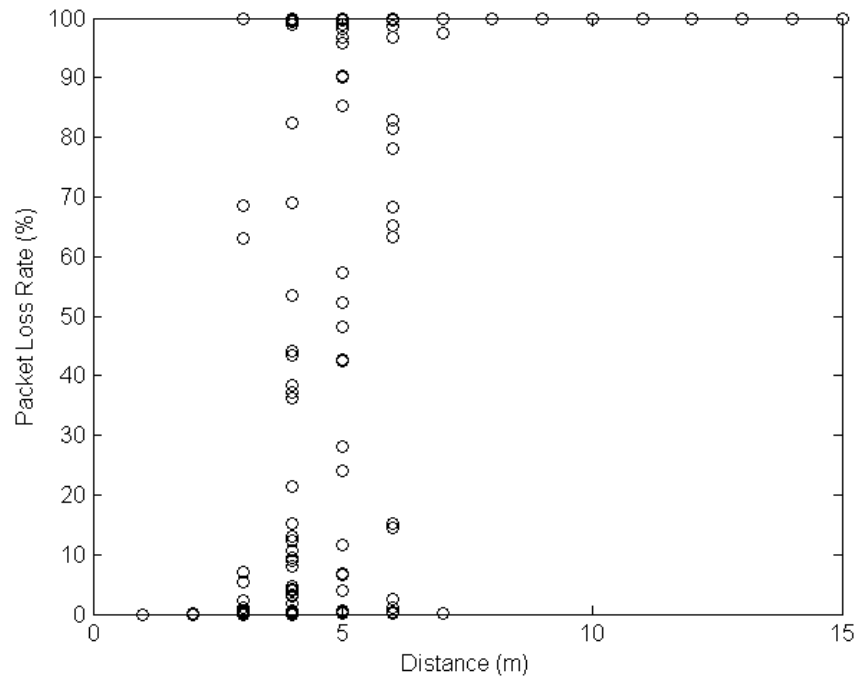

**Fig. 6.** Packet loss rate with respect to distance (transmission power: -10dBm)

In summary, the PLR of wireless channels could vary significantly with respect to distance and transmission power. The radio range and the size of connected region (corresponding to reliable links) depend heavily on the transmission power. Generally speaking, they increase with the level of transmission power. In the transitional region, the wireless link becomes unreliable, featuring uncertain PLR. In the context of cyber-physical control, this unreliability of wireless links and the uncertainty of PLR raise crucial challenges.

## 6. Packet Loss Compensation

In this section, we present a simple solution that can be used in cyber-physical control systems over WSANs to cope with packet loss. Due to the above-analyzed characteristics of wireless channels, it is important to develop a platform-independent paradigm to enhance the reliability of WSANs under lossy conditions. A desirable solution should be widely applicable to diverse application scenarios with different system and environment setups.

Based on these observations, we attempt to develop an approach to packet loss compensation which conforms to the following principles: 1) to modify only the application layer of the networks without exploiting any application-specific (lower layer) network protocols, 2) not to use any statistic information about the distribution of packet loss rate in any specific WSAN, and 3) not to use the knowledge about the models of the controlled physical systems and the controller design of the applications.

### 6.1. A Simple Solution

In this work, we employ a simple method on the actuator nodes to cope with packet loss occurring in WSANs. The basic idea is: whenever a sensor data packet is lost, the actuator will produce an estimate of the sensed value and compute the control command (usually called control input in control terms) based on this value. Note that in our previous work [3], we proposed a method that predicts directly the control command based on previous control command values, which is different from the method used here.

Let *y* denote the controlled variable, i.e. measurement of system output. Suppose that the *k*-th sensed data, i.e. *y(k)*, is lost. From a control perspective, *k* corresponds to sampling instance in discrete time. The actuator will calculate an estimate of *y(k)*, denoted $\hat{y}(k)$, using a predication algorithm, say $f(\bullet)$. Accordingly, we have:

$$\hat{y}(k) = f(y(k-1), y(k-2), ..., y(k-m)) \qquad (1)$$

where *m* represents the number of history data that are stored temporarily for the purpose of prediction. Intuitively, *m* is an important design parameter of the algorithm, which determines the overhead in terms of memory requirements.

For simplicity, we do not consider the effect of delay. More specifically, we assume zero transmission delay in this work. Using Equation (1), the actuator predicts *y(k)* based on the previous *m* consecutive measurements (which are also possibly predicted values) in the case of packet loss. $\hat{y}(k)$ will then be used to compute the control command. Given that the accuracy of the prediction is sufficiently high, proper actions will be performed on the controlled physical system regardless of the loss of the sensed data. In this way, the effect of packet loss on the performance of the control applications can be substantially reduced. From the application's viewpoint, the reliability of the WSAN is improved. It is worth noting that the value of $\hat{y}(k)$ will be stored (as *y(k)*) when *y(k)* is lost, and this value will then be used as *y(k)* whenever necessary (e.g., if a later packet is lost).

The work flow of the actuator (running at every sampling instance) can be illustrated as follows:
  *Input*: Sensed data
  *Output*: Control command
  *Begin*
    If the sensed data *y(k)* is lost then
      Compute $\hat{y}(k)$ using (1)
      Set $y(k) = \hat{y}(k)$

End if
    Produce control command with respect to *y(k)* (through executing control algorithm)
    Store *y(k)* into memory
    Discard *y(k–m)* in the memory
    Perform actions corresponding to the control command
  **End**

It is clear that this solution is quite simple. The major overhead is a small fraction of memory for temporarily storing the previous *m* measurements. Despite this, it does not depend on any knowledge about the underlying platform, environment, link quality characteristics, models of the controlled systems, or controller design.

### 6.2. Prediction Algorithms

It is intuitive that the performance of the above solution is closely related to the prediction accuracy of the algorithm employed, i.e. $f(\bullet)$. Therefore, the design/choice of the prediction algorithm is important in this context. Furthermore, resource-constrained sensor and actuator nodes favor simple algorithms that yield small computational overheads.

Many existing prediction algorithms could be employed here. In this work we explore three types of classic prediction algorithms, which are detailed below.

*Algorithm 1*

This algorithm is based on the assumption that the state of the physical system does not change during the last sampling period. It can be formulated as follows:

$$\hat{y}(k) = y(k-1) \qquad (2)$$

*Algorithm 2*

The second algorithm computes a moving average of the previous *m* samples. This average is then used as the predicted value. Accordingly, we have:

$$\hat{y}(k) = \frac{1}{m}\sum_{i=1}^{m} y(k-i) \qquad (3)$$

*Algorithm 3*

A property of Algorithm 2 is that it treats every previous measurement equally. Algorithm 3 represents an alternative method by giving different weights to previous measurements. More specifically, this algorithm is given by:

$$\hat{y}(k) = \alpha \times y(k-1) + (1-\alpha) \times y(k-2) \qquad (4)$$

where $\alpha$ is a design parameter that commonly satisfies $0 < \alpha < 1$.

## 7. Simulation Results

In this section, we conduct simulations using Matlab to evaluate the performance of the solution and algorithms presented in the previous section. The objective is to examine their potential in coping with packet loss in cyber-physical control systems over WSANs. We first describe the simulation settings, and then analyze the simulation results.

### 7.1. Setup Overview

Consider a physical system that can be modeled in transfer function as follows:

$$G(s) = \frac{1000}{s^2 + s} \tag{5}$$

The controller uses the PID (proportional-integral-derivative) control law, the most popular control law in practical control applications. Controller parameters are chosen according to [26]. The sampling period of the sensor is set to 10ms. The integral of absolute error (IAE, a widely-used performance metric in control community) is recorded to measure the performance of the control application. IAE is calculated as:

$$IAE(t) = \int_0^t |r(\tau) - y(\tau)| d\tau \tag{6}$$

where $t$ denotes (simulation) time and $r(t)$ the desired system output. Note that, in general, larger IAE values imply worse performance.

To examine the effects of different levels of packet loss, we consider the following values of packet loss rates: 0, 20%, and 40%, respectively. To reflect the statistical effect of random packet loss, each simulation runs 100s, which is equal to ten thousand sampling periods. For Algorithm 2, we set $m$ to 3. For Algorithm 3, $\alpha = 0.7$.

### 7.2. Results and Analysis

Fig. 7 depicts a piece of the system output when there is no packet loss (i.e. PLR = 0). Although there is no need of loss recovery, this case will serve as a baseline for studying the impact of packet loss. It can be seen that the control performance is quite good when no packet is lost (i.e. all packets are successfully transmitted).

When the PLR becomes 20%, the (accumulated) IAE values corresponding to the three algorithms presented in Section 6.2 are given in Fig.8, along with the IAE value in the case of no compensation (denoted NON). In addition, the IAE value for the case of no packet loss (denoted NOLOSS) is also given for the purpose of comparison. As we can see, when the PLR changes from zero to 20%, the IAE value increases from 7.1 to 32.9 if no compensation method for packet loss is employed. This confirms to the fact that packet loss deteriorates control performance. In this case, the three algorithms yield comparable control performance which makes insignificant difference from the case of no compensation.

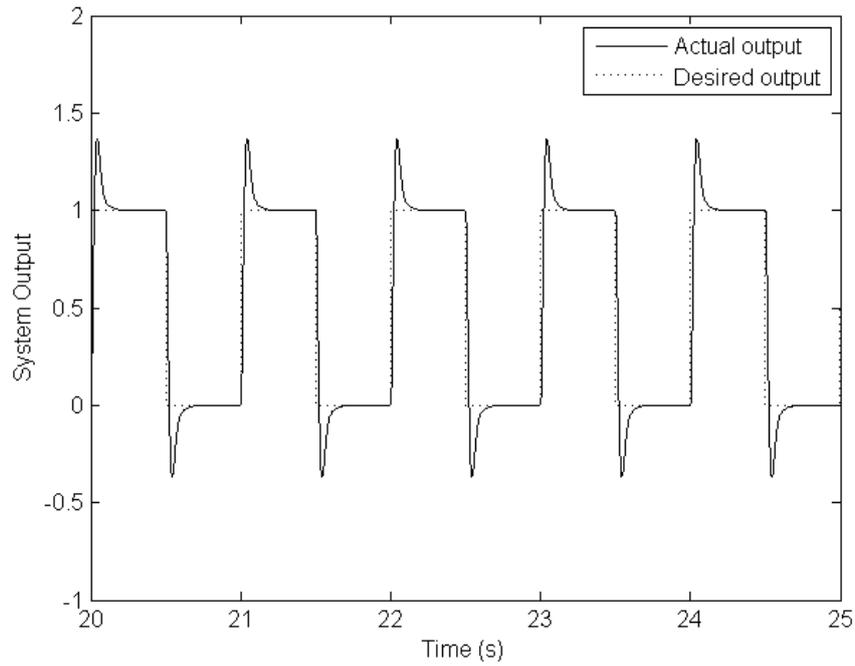

**Fig. 7.** System output in the case of no packet loss

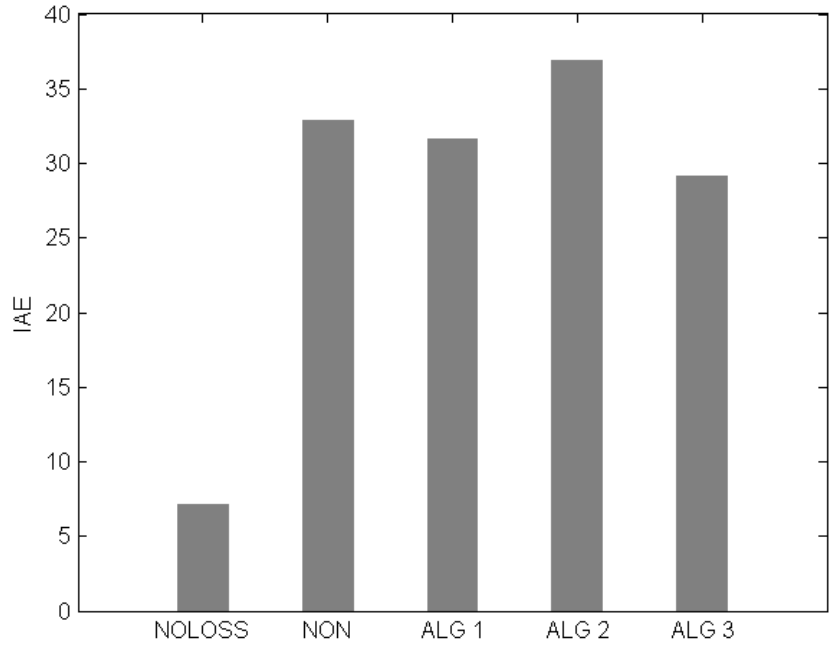

**Fig. 8.** Accumulated IAE values with a PLR of 20%

Fig.9 depicts the accumulated IAE values associated with the case in which PLR is 40%. Due to increase in PLR, the IAE values in this case are much larger than those in Fig.8, implying that the control performance become worse. For example, for the solution that does not use any compensation method, the IAE value reaches 169.6 when PLR is 40%, while this value is 32.9 with a PLR of 20%. In the case of a PLR of 40%, all of the three algorithms result in better control performance than the no compensation solution, which is indicated by the relatively smaller IAE values. For instance, the IAE value associated with Algorithm 3 is 100.6, which is less than 60% of that of the no compensation solution.

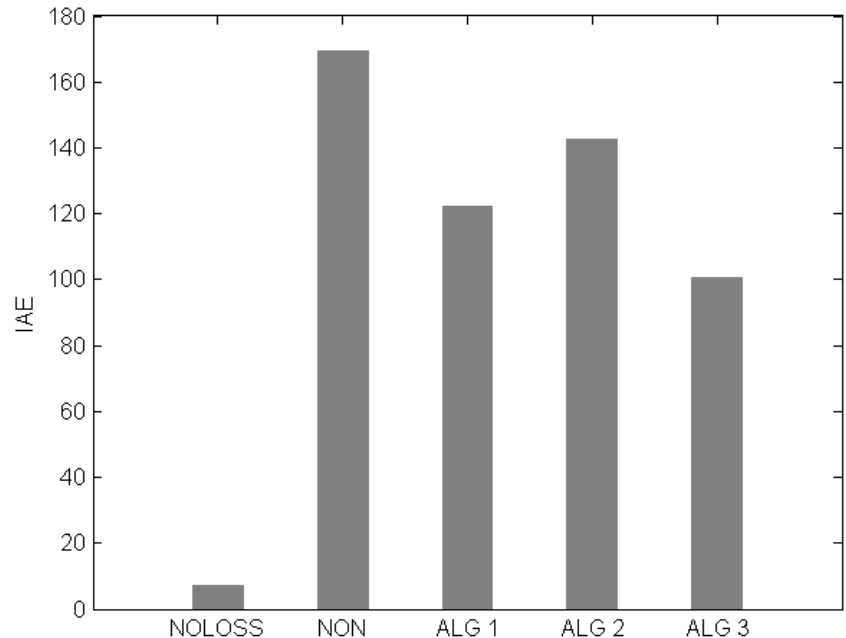

**Fig. 9.** Accumulated IAE values with a PLR of 40%

It is noteworthy that different results might possibly be obtained for different simulation runs due to the uncertainty of packet loss, though every simulation run in our experimentation lasts a considerably large number of (i.e. ten thousand) sampling periods. Figs. 8 and 9 are only some results that are representative of many other results we have obtained. From these results we could make the remark that the prediction algorithm needs to be designed very carefully in order to effectively tackle the problem of unpredictable packet loss, and, furthermore, this is often difficult.

## 8. Conclusion

This chapter has dealt with the topic of how to construct cyber-physical control systems over WSANs that are unreliable. We have examined the system architecture and relevant QoS challenges. The behavior of wireless channels in terms of packet loss rate has been captured by means of simulations using a realistic link-layer model. We presented a simple solution for addressing the problem of packet loss. This solution can be used as a generic framework in which many existing prediction algorithms are applicable. We have also conducted simulations to evaluate the performance of three simple algorithms. The results give some interesting insights useful for control over lossy WSANs. However, it remains open to devise simple yet efficient prediction algorithms for packet loss compensation.

## Acknowledgments

This work is partially supported by Natural Science Foundation of China under Grant No. 60903153.